\documentclass[twocolumn,10pt,prl]{revtex4-1}
\usepackage[utf8]{inputenc}

\def\mysections#1{{\bf #1.} } 
\usepackage{hyperref}
\usepackage[dvips]{graphicx}
\usepackage{epsfig,amsmath,amssymb,verbatim,mathrsfs,array,layout,textcomp,amssymb,latexsym,slashed,graphicx,booktabs,color,mathtools}

\newcommand{\beq}{\begin{equation}}
\newcommand{\eeq}{\end{equation}}

\def\beqa{\begin{eqnarray}}
\def\eeqa{\end{eqnarray}}
\def\bea{\begin{eqnarray}}
\def\eea{\end{eqnarray}}
\newcommand{\no}{\nonumber}
\newcommand{\bv}{\left(\begin{array}{c}}
\newcommand{\ev}{\end{array}\right)}
\newcommand{\bmtwo}{\left(\begin{array}{cc}}
\newcommand{\bmthree}{\left(\begin{array}{ccc}}
\newcommand{\emn}{\end{array}\right)}
\newcommand{\bmtwoc}{\left\{\begin{array}{cc}}
\newcommand{\bmthreec}{\left\{\begin{array}{ccc}}
\newcommand{\emnc}{\end{array}\right\}}
\newcommand{\ba}{\begin{array}}
\newcommand{\ea}{\end{array}}

\newcommand{\GeV}{\text{ GeV}}

\def\lsim{\mathrel{\rlap{\lower4pt\hbox{\hskip1pt$\sim$}}
     \raise1pt\hbox{$<$}}}         
\def\gsim{\mathrel{\rlap{\lower4pt\hbox{\hskip1pt$\sim$}}
     \raise1pt\hbox{$>$}}}         

\begin{document}
\font\mini=cmr10 at 0.8pt

\title{
Inflating to the Weak Scale
}

\author{Michael Geller${}^{1,2}$}\email{mic.geller@gmail.com }
\author{Yonit Hochberg${}^{3}$}\email{yonit.hochberg@mail.huji.ac.il}
\author{Eric Kuflik${}^{3}$}\email{eric.kuflik@mail.huji.ac.il}

\affiliation{${}^1$Maryland Center for Fundamental Physics, Department of Physics, University of Maryland, College Park, MD 20742, USA}
\affiliation{${}^2$Department of Physics, Tel Aviv University, Tel Aviv, Israel}
\affiliation{${}^3$Racah Institute of Physics, Hebrew University of Jerusalem, Jerusalem 91904, Israel}

\begin{abstract}
We present a new solution to the hierarchy problem, where the Higgs mass is at its observed electroweak value because such a patch inflates the most in the early universe. If the Higgs mass depends on a field undergoing quantum fluctuations during inflation, then inflation will fill the  universe with the Higgs mass that corresponds to the largest vacuum energy. The hierarchy problem is solved if the maximum vacuum energy occurs for the observed Higgs mass. We demonstrate this notion with a proof-of-principle model containing  an axion, a modulus field and the Higgs, and show that inflation can be responsible for the weak scale. 
\end{abstract}

\maketitle

\section{Introduction}

The hierarchy problem between the electroweak (EW) and Planck scales has been a driving force in particle physics for many decades. Popular solutions along the years have included the introduction of new particles at or close to the weak scale via supersymmetry, extra dimensions  and composite Higgs. Recent years have sparked new directions for addressing the hierarchy problem, including ideas such as Refs.~\cite{Chacko:2005pe,Graham:2015cka,Giudice:2016yja,Arkani-Hamed:2016rle,Cheng:2018gvu,Hook:2018jle,Cohen:2018mgv}.  

Here we propose an alternative solution to the hierarchy problem, where inflation can be responsible for the weak scale permeating the entire universe.

In the early universe, a scalar field can undergo quantum fluctuations, leading to a distribution in space of  the field values. Each patch will have a different vacuum energy, depending on the field value in that volume. The patches with the largest vacuum energy will inflate the fastest, and by the end of inflation, most of the universe will have the corresponding field value. If the Higgs mass parameter is dynamical, and depends on the fluctuating field, then at the end of inflation the universe will be filled nearly entirely with the single Higgs mass that maximizes the scalar potential. The hierarchy problem will then be solved if the maximum vacuum energy occurs for the observed Higgs mass, and the inflationary period  is long enough to fill the entirety of space with the observed value.

This letter is organized as follows. We start by elaborating on the basic concept of the proposed mechanism. To provide proof-of-concept, we then present a concrete model which realizes this solution to the naturalness problem. We describe the conditions under which the model is viable---namely, that the observed Higgs mass spans the entire universe, without eternal inflation or fine tuning.  We conclude with some discussion on further directions and improvements.

\section{Basic Concept}
During inflation every scalar field has an uncertainty of the order of the inflationary Hubble scale. Assuming the potential of the  field is very flat, classical rolling down this potential is negligible. However, the field does undergo a random walk due to the quantum uncertainty, generating a spread in field value that increases with time. At any given time, each patch of the universe will then have a different field value and correspondingly a different vacuum energy. The patches with the highest vacuum energy will inflate the most, and if inflation is long enough, the corresponding field value will fill out most of space by the end of inflation. 

If such a scalar field is coupled to the Higgs boson, then the Higgs mass parameter in the universe will correspond to the value of the scalar field at the maximum of the potential. To solve the hierarchy problem, the scalar potential is constructed such that the maximal vacuum energy occurs when the Higgs mass fits the observed value in our universe. There might be more than one way to build such a model. In our proof-of-concept model, given in detail in the next sections, we single out the correct EW mass parameter by using the QCD axion which is sensitive to the Higgs mass. This axion has a clockwork-like potential~\cite{Choi:2014rja,Kaplan:2015fuy,Choi:2015fiu}, which was  used in Ref.~\cite{Graham:2015cka} to scan the Higgs mass parameter.

\section{Model}

Here we construct an explicit model that realizes our proposed mechanism to solve the hierarchy problem via inflation. This proof-of-concept model  serves to illustrate that the proposed mechanism can be realized. The scalar fields in the model include the SM Higgs boson $h$, the QCD axion $a$,  a modulus field $\phi$ and the inflaton. We are agnostic about the inflaton sector, which is only assumed to end inflation after a long inflationary period. The potential of the modulus and axion fields is taken to be 
\bea
V &=& \left( M^2 + y M \phi + \ldots \right) h^2 +\lambda h^4 + y M^3 \phi  + \ldots \no\\
   &&+ \frac{a}{f} G\tilde{G} + \Lambda_{H}^4 \cos \frac{a}{F}\,.
\label{eq:potential}
\eea
Here $M$ is the cutoff of the theory---the `natural' size of the Higgs mass---and $G$ is the gluon field strength. The axion potential $\Lambda_{H}^4 \cos \frac{a}{F}$ is generated  by the non-perturbative
dynamics of a hidden confining gauge theory,  with confinement scale $\sim\Lambda_{H}$. The $+\dots$ terms refer to terms that are higher order in $(y \phi)$. The parameter $y$ is a spurion that breaks a shift symmetry for $\phi$ (which will be very small), and thus the potential above is technically natural.  As we will see later, the field $\phi$ can be a compact field; the term modulus is used loosely here.

The Higgs mass parameter  in Eq.~\eqref{eq:potential} is promoted to a dynamical field,
\beq
\mu(\phi)^2 = M^2 + y M \phi+ \ldots 
\eeq 
The $\phi$ field begins 
at a large negative value ${\phi \lsim -M/y} $, where the Higgs mass parameter is large and negative, ${\mu^2 \lsim -M^2}$.  The field diffuses to larger values where $\phi \sim -M/y$, including where the two contributions to the Higgs mass parameter are `tuned' to give the observed value ${\mu_{\rm obs}^2 \simeq -(125./\sqrt{2})^2 \GeV^2}$. Below we will see how this point is `naturally' picked out.
Since the field $\phi$ fluctuates, and these fluctuations drive the mechanism, we call it the {\it fluctuon}. 
We note that, similar to Ref.~\cite{Graham:2015cka},  since $y$ will be a small coupling, the field $\phi$ undergoes excursions much larger than the cutoff and the Planck scale. We expect that this should not necessarily pose a problem for a UV-completion of this model, {\it e.g.} within a clockwork framework.

\begin{figure}[t!]
\includegraphics[width=.48\textwidth]{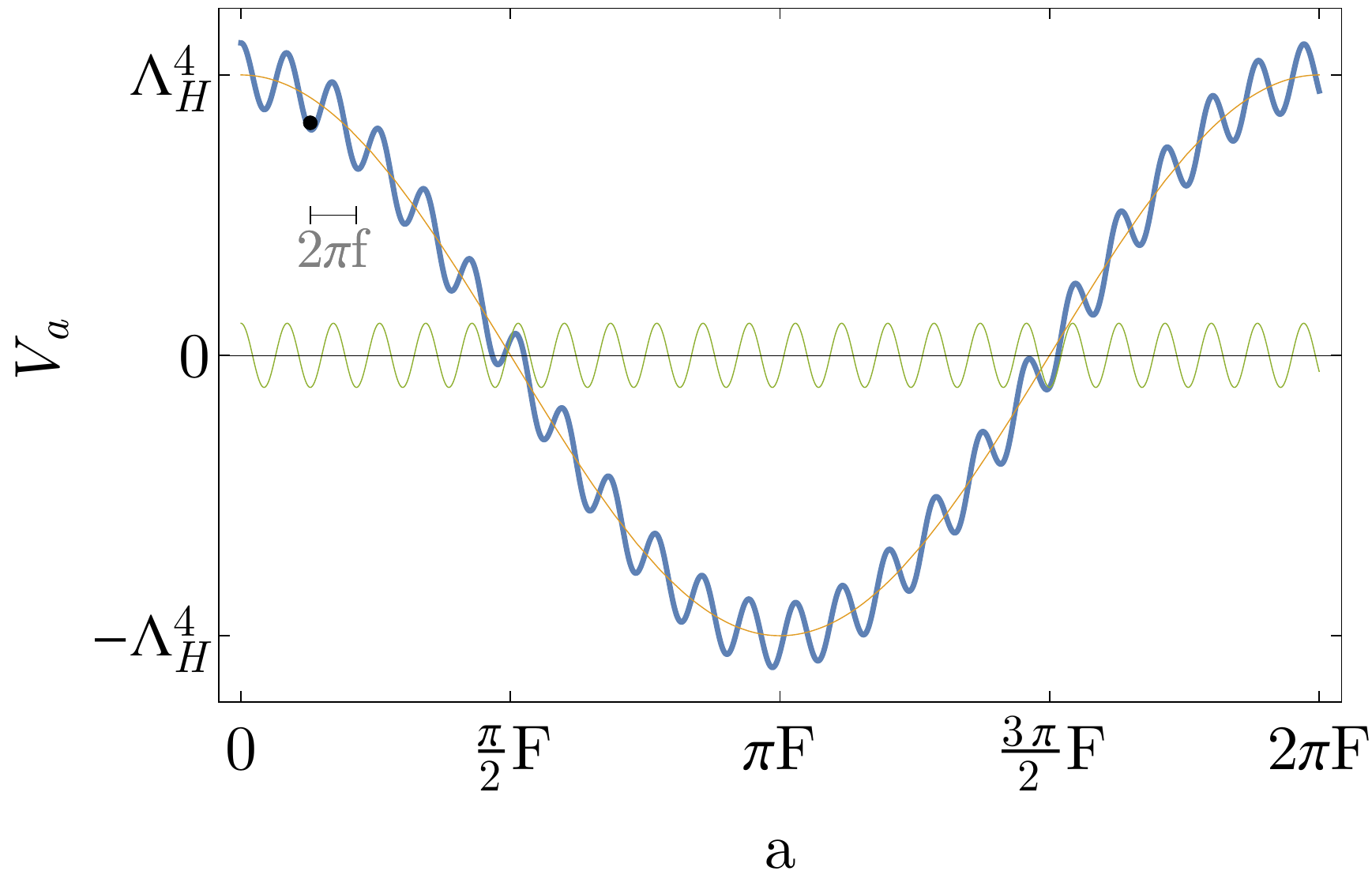} \\\includegraphics[width=.48\textwidth]{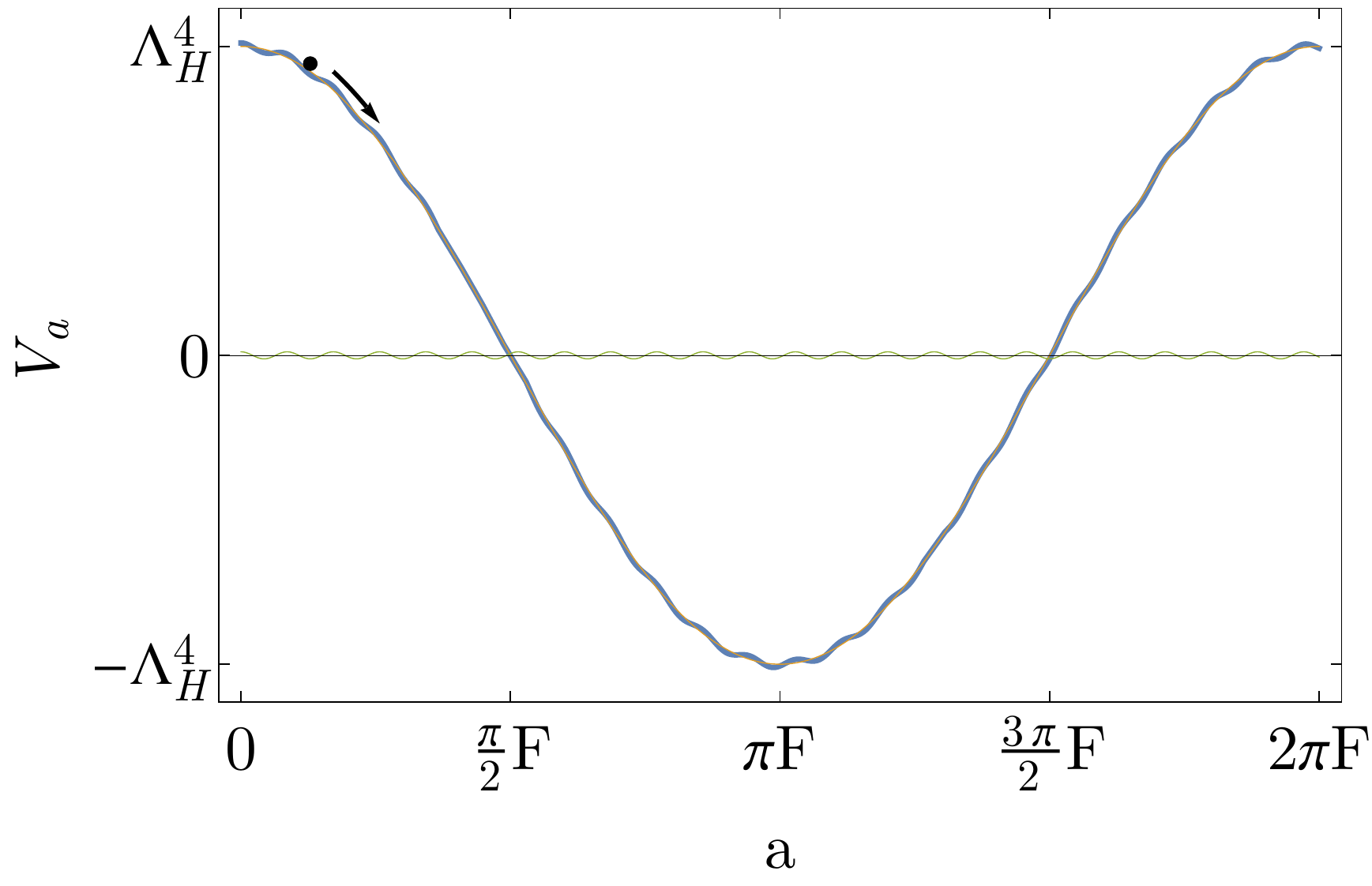}
\caption{\label{fig:Va}
The axion potential. {\it Top:}  The axion is initially misaligned from the global minimum and is trapped at a local minimun of the potential. {\it Bottom:} As $\phi$ increases and the Higgs VEV drops, the QCD-induced part of the potential releases its barriers and the axion is free to roll down to the global minimum. 
}
\end{figure}

\begin{figure}[t]
\includegraphics[width=.48\textwidth]{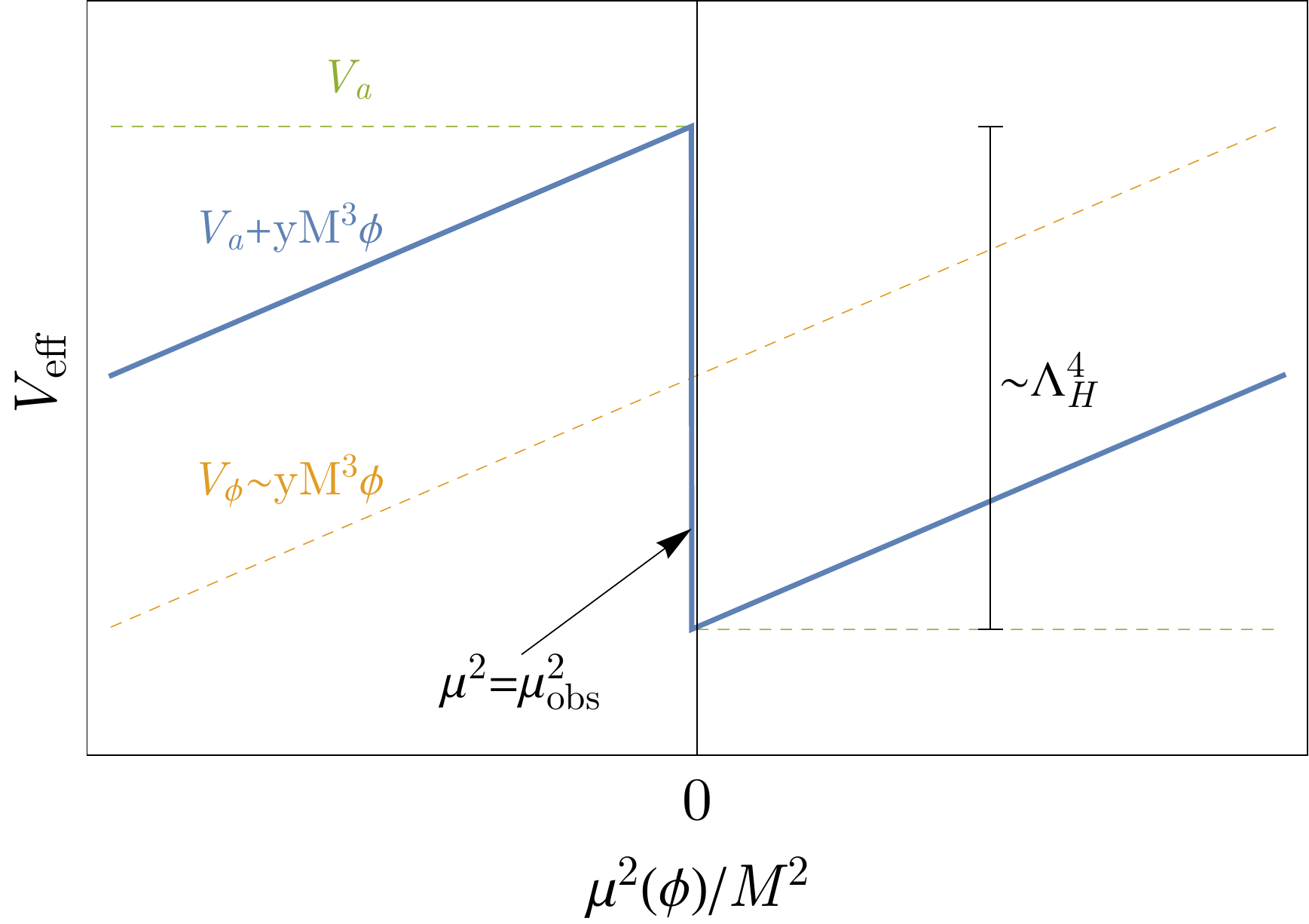} 
\caption{\label{fig:H}
Total contributions to the vacuum energy. The green dashed curve shows the axion potential energy at the end of inflation; the dashed orange curve depicts the $\phi$ potential; and the total contributions are shown in solid blue.
}
\end{figure}

The axion potential can be seen in Fig~\ref{fig:Va}. It is comprised of the addition of two modulating potentials, one with large amplitude and period (from the confining hidden sector) and one with small  amplitude and period (from the QCD sector), with $F \gg f$,
\beq
V_a = \Lambda(\phi)^4 \cos {a/f} + \Lambda_H^4 \cos{a/F}\,.
\eeq
 The fluctuon field $\phi$ begins at a large negative Higgs mass, and QCD dynamics generates a potential for the axion, $\Lambda(\phi) \sim \Lambda_{\rm QCD}$. Initially, the axion is misaligned from the global minimum, trapped at a local minimum of the QCD potential (see top panel). Near the observed EW VEV, the amplitude, $\Lambda(\phi)^4$,  is approximately linear in the Higgs VEV, and therefore varies  with $\phi$. For larger $\phi$ values (a consequence of diffusion), the Higgs VEV decreases, and the barriers created from the QCD potential falls. If the slope of the barriers of the QCD potential drops below the slope of the potential from the hidden confining sector, the axion will be free to roll down the potential (see bottom panel). The axion parameters are chosen such that the barriers disappear when the Higgs mass parameter is smaller (or equal) to the observed value in our universe, {\it i.e.}
\begin{equation}
\frac{\Lambda_H^4}{F} \sim \frac{\Lambda^4_{\rm QCD}}{f}. \label{eq:lambdas}
\end{equation}

The potential energy (from $\phi$ and $a$) contributing to the Hubble scale can be separated into two contributions: the linear potential from the fluctuon $\phi$, and the potential energy of the axion when it has reached it's local minimum. Integrating out the axion contribution and writing the potential in terms of the Higgs mass parameter gives
\beq
V_{\rm eff} \simeq M^2 \mu(\phi)^2 -2 \Lambda_H^4 \Theta\left(\mu(\phi)^2  - \mu^2_{\rm obs}\right)\,.
\label{Veff}
\eeq
We plot this effective potential in Fig.~\ref{fig:H}. For smaller $\phi$, the potential increases as $V_\phi \sim y M^3 \phi \sim M^2 \mu^2$, but there is also a constant contribution from the axion which is trapped in a barrier near the top of the potential  $V_a \sim \Lambda_H^4$. However, as $\phi$ increases enough, the barrier falls, and the axion moves towards the global minimum, $V_a \sim -\Lambda_H^4$. The energy density is therefore maximized when the Higgs VEV is at the measured value in our universe, which is much less than the cutoff of the theory. 

The potential described in Fig.~\ref{fig:H} can potentially have larger vacuum energies for much larger positive values of $\mu^2({\phi})$ than depicted. 
A necessary requirement is that the confinement scale of the hidden confining gauge theory be larger than the cutoff,
\beq
\Lambda_H > M,
\eeq
so that the drop in the vacuum energy is significant at the observed Higgs mass. Nevertheless, for the linear potential in Eq.~\eqref{Veff}, the potential energy will eventually exceed the potential energy at $\mu^2_{\rm obs}$.
Thus a requirement of the model is that the potential of $\phi$ have a global maximum for $|\phi |\lesssim {M}/{y}$, such that the maximum of the total potential stays at $\mu^2_{\rm obs}$. One possibility is that $\phi$ is compact with a sinusoidal potential and an amplitude $\lsim\Lambda^4_H$. Alternatively, since the full potential of $\phi$  contains terms such as $y^2 M^2 \phi^2$, $y^3 M \phi^3$ etc., it  could very well be that  $V(\phi)$ is maximized on the range $|\phi |\lesssim {M}/{y}$, without a periodic potential.

The patches where the Higgs VEV matches  the observed value have the highest vacuum energy, and therefore expand exponentially faster than the rest of the universe. If inflation is long enough, most of the universe (by volume) will have the observed Higgs mass. Thus, inflation solves the hierarchy problem  
by filling the universe with the correct Higgs mass.

\section{Basic Requirements}

Having set up the model, we now move to the conditions such that the mechanism works within this model. We derive the necessary conditions on the model parameters such that the universe is filled with $\mathcal{O}({\rm EW})$ scale Higgs mass, and that the probability of being found in a patch with Higgs mass much larger than the observed value is close to zero. Furthermore, we will give the conditions that the spread in the observable universe is small. 

We assume that the inflaton dynamics is external to the dynamics of the Higgs mass, and that the inflaton dominates the energy density during inflation. The Hubble scale at this time can be expanded around the $\phi/a$ contribution,
 \beq
 H \simeq H_{\rm inf} + \Delta H, ~~~ \Delta H \simeq  \frac{V}{2 H_{\rm inf} m_{\rm pl}^2}
 \label{eq:deltah}
 \eeq
with $V= y M^3 \phi +V_a$, where $V_a$ is the vacuum energy of the axion and $H_{\rm inf} \equiv M_{\rm inf}^2/m_{\rm pl}$ is the Hubble scale from inflation. Note that since the inflaton dominates the energy density, $ M_{\rm inf} \gtrsim M$.

First, it is important that when the barriers release the axion, it rolls sufficiently down the potential so as to generate a sizable change in the vacuum energy of the universe. This imposes two requirements: {\it (i)} that the classical motion of 
the axion dominates over its quantum fluctuations, in order that all patches roll towards the minimum; and {\it (ii)}, that during inflation the axion has  rolled enough such that its contribution to the potential has dropped relative to the variance in the $\phi$ potential.

In one Hubble time, the axion field undergoes quantum fluctuations of order $\Delta a_{\rm qu} \approx H_{\rm inf}$, and rolls classically $\Delta a_{\rm cl}\approx \frac{1}{H_{\rm inf}^2}  \frac{\partial V(a)}{\partial a}  $. For the potential Eq.~\eqref{eq:potential}, the requirement  of classical motion beating quantum fluctuations for the axion thus implies an upper bound on the cutoff of the theory:
\beq\label{eq:M}
M \lesssim  M_{\rm inf} < \frac{\Lambda^{2/3}_{\rm QCD} m^{1/2}_{\rm pl} }{f^{1/6}}  \simeq 10^7~{\rm GeV} \left( \frac{10^8 ~\rm GeV}{f} \right)^{1/6}\!\!\!\!\!.
\eeq
Here we used the relationship in Eq.~\eqref{eq:lambdas}, that ${V_{\phi,a}< V_{\rm inf}}$, and took $\Lambda_{\rm QCD}=0.1$~GeV. Translating this bound to a bound on $F$, gives $F > 10^{40}$~GeV, for the values taken in Eq.~\eqref{eq:M}, which can be accommodated in a clockwork mechanism.

 Further requiring condition {\it (ii)}, that the axion contribution to the potential decreases by $\mathcal{O}(\Lambda_H^4)$, by moving $\mathcal{O}(F)$,  during inflation, imposes that inflation takes place for a sufficiently long time, 
 \begin{equation}\label{eq:Ninf}
N > \frac{M^4 M_{\rm inf}^4 f^2}{\Lambda^8_{\rm QCD} m_{\rm pl}^2} \simeq10^{44}   \left( \frac{f}{10^8 ~\rm GeV} \right)^{2} \left( \frac{M M_{\rm inf} }{(10^7 ~{\rm GeV})^2} \right)^{4}\, ,
 \end{equation}
where $N=H_{\rm inf} t_{\rm inf} $ is the number of e-folds and $t_{\rm inf}$ is the time that inflation ends. In the above we assume the initial axion misalignment does not place the axion at the top of the potential. However, if the distribution of the initial condition for the axion is spread over local minimum, then the axion will be at the top of the potential in most of the Hubble patches. We have checked that our mechanism works in that case as well and that the requirement in Eq.~\eqref{eq:Ninf} is then weakened. We leave this for future work.
 
Next we study the diffusion and growth of the distribution of $\phi$ during inflation, which is similar to the evolution originally studied in models of chaotic eternal inflation~\cite{Linde:1986fd,Goncharov:1987ir}, the Higgs field during inflation~\cite{Espinosa:2007qp,Kobakhidze:2013tn,Hook:2014uia}, and the relaxion~\cite{Nelson:2017cfv,Gupta:2018wif}. One needs to find the fraction of the physical volume of the universe that is filled with the observed EW scale, and require that this probably is very close to unity. A complete analysis of the probability distribution requires solving the full Fokker-Planck diffusion equations for $\phi$ and $a$, which is being studied in detail in upcoming work~\cite{future}. 

For the case where quantum fluctuations of $\phi$  dominate over classical motion (which we will quantify below) and that the fluctuon $\phi$ contribution to the Hubble scale is small, an approximate solution can be obtained.  The distribution factorizes into two contributions: the gaussian distribution that characterizes standard diffusion in a comoving volume, and the growth of the volume with the Hubble scale:
 \bea\label{eq:P}
P(\phi,t) &\sim& P_{\rm diff}\times P_{\rm growth}\\
&\sim& \exp \left( -\frac{\left[\phi(t) - \phi_{\rm cl}(t)\right]^2}{H^3_{\rm inf} t}\right)  \times \exp \left( 3 \Delta H\, t\right) \,, \label{eq:pdiff} \nonumber
 \eea
 where $\phi_{\rm cl}(t)$ is the classical expectation value of the fluctuon field, and $\Delta H$ is given in Eq.~\eqref{eq:deltah}. 
 
 In order for the fluctuon to move up the potential, its quantum fluctuations must be larger than the classical motion, which is condition {\it (iii)}: 
  \beq \label{eq:yquant}
 y < \frac{M_{\rm inf}^6}{M^3 m_{\rm pl}^3}   \simeq 10^{-33} \left( \frac{M_{\rm inf}}{10^7 ~\rm GeV} \right)^6  \left( \frac{10^7 ~\rm GeV}{M} \right)^3 \,.
\eeq
 The growth from inflation has dominated over the diffusion when the EW vacuum has a larger density than the classical value, {\it e.g.}, $  P(\phi_{\rm EW},t_{\rm inf})  \gg   P(\phi_{\rm cl},t_{\rm inf}) $. Requiring inflation is long enough---condition~{\it (iv)}--- imposes that the number of e-folds must be greater than
  \beq\label{eq:Nefold}
 N > \frac{m_{\rm pl}}{y M } \simeq  10^{44}   \left( \frac{10^7 ~\rm GeV} {M}\right) \left( \frac{10^{-33}} {y}\right) ,
 \eeq
  where we used the fact that $\phi_{\rm EW} - \phi_{\rm cl}(t_{\rm inf}) \approx M/y$.

 We now require that the EW-scale VEVs are present across the entire universe, so that the hierarchy problem is solved everywhere. 
After a long enough time of inflation (given by Eq.~\eqref{eq:Nefold}), the  distribution is dominated by inflationary growth, $ P(\phi,t_{\rm inf}) \propto  \exp\left(3 \Delta H\, t_{\rm inf}\right) \Theta(\phi(t_{\rm inf})-\phi_{\rm EW})$.   
 The probability to be found in a patch less than $-\delta \mu^2$ from the observed value is 
 \beq
P \left(\mu^2<\mu^{ 2}_{\rm obs}-\delta \mu^2\right) \simeq \exp \left[-\frac{ 3M^2 t_{\rm inf}}{H_{\rm inf} m_{\rm pl}^2} \delta \mu^2 \right] \,.
 \eeq
Requiring this probability be small for $\delta \mu^2 \sim |\mu_{\rm obs}|^2$, which is condition~{\it (v)}, imposes
\begin{equation}\label{eq:Ndelmu}
N > \frac{H_{\rm inf}^2 m_{\rm pl}^2}{M^2 |\mu_{\rm obs}|^2} \simeq 10^{10} \left( \frac{M_{\rm inf}} {10^7 ~\rm GeV}\right)^4    \left( \frac{10^7 ~\rm GeV}{M} \right)^2\,.
\end{equation}
Given Eqs.~\eqref{eq:Nefold} and \eqref{eq:Ndelmu}, the likelihood of a patch with Higgs mass much larger than the observed value is close to nil. 

Therefore, given moderately long inflation, the exponential growth of the inflationary period will fill the universe with EW-scale VEVs,  that are much smaller than the cutoff. The longer the period of inflation, the smaller the spread in the electroweak scale is across the universe.  In the following section, we will discuss post-inflationary dynamics of the fluctuon field and the spread of the EW scale within the observable universe.

 \section{Post-inflationary dynamics}

If the fluctuon field $\phi$ continues to classically roll after inflation,  it can change the electroweak scale and the vacuum energy that was picked out by inflation. Thus we require that the change in the Higgs mass and the vacuum energy from  post-inflationary classical rolling of $\phi$ be less than the observed values today. The post inflationary change in $\phi$ until today can be estimated as
$
  \Delta \phi_0 \approx {y M^3}/{H_0^2},
$
where $H_0$ is the Hubble scale today. The subsequent constraints on the slope from the Higgs mass and vacuum energy, conditions~{\it (vi)} and {\it(vii)}, are
\beq\label{eq:ymH}
y < \frac{|\mu| H_0}{M^2} \simeq 10^{-54}   \left( \frac{10^7 ~\rm GeV} {M}\right)^{2} 
\eeq
and 
\beq\label{eq:yvac}
y < \frac{\Lambda_{cc} m_{\rm pl}}{M^3} \simeq 10^{-86}  \left( \frac{10^7 ~\rm GeV} {M}\right)^{3}\, ,
\eeq
respectively, where  
$\Lambda_{cc}\simeq10^{-122} m_{\rm pl}^2$ is the observed cosmological constant. Using this constraint, along with Eq.~\eqref{eq:Nefold}, we find that in order to avoid rolling of the EW scale after inflation, the number of e-folds must be $N > 10^{97}$ for the parameters chosen.   

Typical models of slow-roll inflation can maximally inflate for 
\beq
N_{\rm slow-roll} \lesssim \frac{m^2_{\rm pl}}{H^2_{\rm inf}} \simeq 10^{44} \left( \frac{10^7 ~\rm GeV} {M_{\rm inf}}\right)^4, \label{infinfl}
\eeq
 before the universe is eternally inflating.  While there are unsolved problems in eternal inflation, it is possible that our universe is eternally inflating, or that the inflationary model is not a standard slow-roll model.  If allowing for eternal inflation, then for small enough $y$ and long enough inflation, the Higgs mass can be driven to arbitrarily uniformity throughout the universe.

 Here we offer an approach which avoids eternal inflation. We take the fluctuon $\phi$ to also be an axion-like field, that develops a periodic potential after inflation ends,
\beq
\Lambda^4_\phi \cos \frac\phi {f_\phi} \,.
\eeq
During inflation this cosine potential is inactive if $H_{\rm inf} >\Lambda_\phi$,
but is reactivated  after inflation. 
For ${\Lambda^4_\phi}/{f_\phi}\gtrsim y M^3$, there are local minima along the $\phi$ potential in which the field can settle. Within quantum field theory~\cite{Gupta:2015uea} and string theory~\cite{McAllister:2016vzi}, the technical naturalness of $y$ for `non-compact axions' can be problematic, but is realizable within the clockwork mechanism~\cite{Choi:2014rja}.

This automatically solves the problem of the fluctuon rolling after inflation, since within each period $f_\phi$, $\phi$ will roll to a fixed value. We  require that in the last $N_{60}=60$ e-folds $\phi$ does not fluctuate larger than $f_\phi$, or else some of the observable patches will end up in different minimum after inflation:
\beq\label{fphi}
 \Delta \phi = \sqrt{N_{60}} H_{\rm inf} < f_\phi .
\eeq
This leads to a condition on $y$: 
\beq \label{eq:ywig}
y < \frac{1}{\sqrt{N_{60}}} \frac{H_{\rm inf}^3}{M^3}   \simeq 10^{-34} \left( \frac{M_{\rm inf}}{10^7 ~\rm GeV} \right)^6  \left( \frac{10^7 ~\rm GeV}{M} \right)^3.
\eeq
Here, eternal inflation can be avoided if the cutoff is brought down to {\it e.g.} $M \lsim 10^6$~GeV.

\section{Model Discussion}

In short, we have derived the necessary conditions on the model parameters for the mechanism to solve the hierarchy problem. During inflation we require that the axion $a$  rolls {\it (i)}~classically [Eq.~(\ref{eq:M})] and {\it (ii)}~sufficiently far down the potential during inflation [Eq.~(\ref{eq:Ninf})], in order that when the barriers release the axion potential energy significantly lowers; that   {\it (iii)}~the fluctuon $\phi$ is dominated by quantum motion so it can move up the potential [Eq.~(\ref{eq:yquant})]; that {\it (iv)}~inflation is long enough so that the fluctuon has spread over a large range and inflationary growth drives the distribution [Eq.~(\ref{eq:Nefold})]; and  {\it (v)}~that the likelihood of a patch with Higgs mass much larger than the observed value is close to nil [Eq.~(\ref{eq:Ndelmu})]. Post-inflation, the fluctuon $\phi$ must not roll sufficiently back down the potential  changing the Higgs mass and vacuum energy, given by conditions {\it (vi)} and {\it (vii)} [Eqs.~\eqref{eq:ymH} and~\eqref{eq:yvac}]. These  last two constraints can be relaxed to Eq.~\eqref{eq:ywig} if $\phi$ generates local minima after inflation.

 We remark that as in Ref.~\cite{Graham:2015cka}, the model above generates a large $\theta$ angle in absence of a tuning, and therefore re-introduces the strong-CP problem that the QCD axion aims to solve. There are several possible solutions to this problem which could be applied to the framework here, such as using a QCD' sector, an additional axion, the Nelson-Barr mechanism or particle production, as in Refs.~\cite{Graham:2015cka,Espinosa:2015eda,Davidi:2017gir,Hook:2016mqo,Tangarife:2017vnd}. These would also relate to the possible signatures of the model.  Additionally, the model proposed here does not address the cosmological constant problem, and we do not explain why the patches with the maximal vacuum energy during inflation will have very little vacuum energy afterwards. 
 We leave a detailed study of expansions of the idea presented here, as well as potential signatures of this mechanism, to future work~\cite{future}.
 
 We leave the constraints from standard cosmology on this mechanism to future work as well~\cite{future}, but we note here that  
 the oscillations of the fluctuon $\phi$ around the minimum after inflation will start early enough, so that $\phi$ will not overclose the universe.  
 
 Finally, we comment on the use of measure probabilities~\cite{Linde:1993nz,Linde:1993xx,GarciaBellido:1993wn,Linde:2010xz} in this work. Since  inflation can be non-eternal, we have adopted the measure to be the physical volume at the end of inflation. Furthermore, the patches with the wrong Higgs VEV will have negative vacuum energy, and will crunch after inflation. Therefore we expect much of the typical ambiguity of defining measures during inflation to be addressable  here.

\section{Summary}

In this work we have proposed a new solution to the hierarchy problem. We have shown that if the Higgs mass is a dynamical variable that is coupled to a fluctuating light field, inflation can be responsible for the electroweak scale by filling the universe with this Higgs mass if it maximizes the scalar potential. We have constructed a simple proof-of-concept model that realizes this idea, and contains a modulus field, an axion, and the Higgs, and derived the conditions for such a model to work. For small enough coupling and long enough inflation, the Higgs mass can be driven to arbitrarily uniformity throughout the universe. We have  shown  that the model is viable without necessarily eternal inflation. There are many possible extensions of this idea and further model realizations, which will be explored in future work.

\mysections{Acknowledgments}
 We are grateful  to Anson Hook, Liam McAllister, Hitoshi Murayama, Gilad Perez and Raman Sundrum  for discussions and comments on the manuscript. 
The work of YH is supported by the Israel Science Foundation (grant No. 1112/17), by the Binational Science Foundation (grant No. 2016155), by the I-CORE Program of the Planning Budgeting Committee (grant No. 1937/12), by the German Israel Foundation (grant No. I-2487-303.7/2017), and  by the Azrieli Foundation. EK is supported by the Israel Science Foundation (grant No. 1111/17), by the Binational Science Foundation  (grant No. 2016153) and by the I-CORE Program of the Planning Budgeting Committee (grant No. 1937/12).  MG was supported in part by the NSF under Grant No. PHY-1620074 and by the Maryland Center for Fundamental Physics (MCFP).  YH and EK would like to express a special thanks to the Mainz Institute for Theoretical Physics (MITP) in Capri for its hospitality and support. The work of MG was in part performed at the Aspen Center for Physics, which is supported by National Science Foundation grant PHY-1607611.

\bibliography{biblioinflate}{}

\end{document}